\documentclass[conference]{IEEEtran}
\IEEEoverridecommandlockouts
\usepackage{cite}
\usepackage{adjustbox}
\usepackage{amsmath,amssymb,amsfonts}
\usepackage{algorithmic}
\usepackage{graphicx}
\usepackage{multicol}
\usepackage{tabularx}
\usepackage{steinmetz}
\usepackage{mathtools}
\usepackage{booktabs}
\usepackage{textcomp}
\usepackage{xcolor}
\usepackage{wrapfig}
\usepackage{subfig}
\usepackage{pifont}
\usepackage{flushend}

\usepackage{mwe}

\usepackage{algorithmic}
\usepackage{algorithm}

\def\BibTeX{{\rm B\kern-.05em{\sc i\kern-.025em b}\kern-.08em
    T\kern-.1667em\lower.7ex\hbox{E}\kern-.125emX}}
    
\begin{document}

\title{
Optimizing FPGA-based Accelerator Design for Large-Scale Molecular Similarity Search 
\\
{\vspace{-10pt}
\large {(Special Session Paper)}
\vspace{-15pt}}
}

\author{\IEEEauthorblockN{
Hongwu Peng\textsuperscript{[1]},
Shiyang Chen\textsuperscript{[2]}, Zhepeng Wang\textsuperscript{[3]}, 
Junhuan Yang\textsuperscript{[4]},
Scott A. Weitze\textsuperscript{[2]},\\
Tong Geng\textsuperscript{[5]},
Ang Li\textsuperscript{[5]}, 
Jinbo Bi\textsuperscript{[1]},
Minghu Song\textsuperscript{[1]},
Weiwen Jiang\textsuperscript{[3]},
Hang Liu\textsuperscript{[2]},
Caiwen Ding\textsuperscript{[1]},}

\IEEEauthorblockA{\textsuperscript{[1]}University of Connecticut, CT, USA.
\textsuperscript{[2]}Stevens Institute of Technology, NJ, USA.
\textsuperscript{[3]}George Mason University, VA, USA. \\
\textsuperscript{[4]}University of New Mexico, NM, USA.
\textsuperscript{[5]}Pacific Northwest National Laboratory, WA, USA. \\
\textsuperscript{[1]}\{hongwu.peng, jinbo.bi, minghu.song, caiwen.ding\}@uconn.edu, 
\textsuperscript{[2]}\{schen94, sweitze, hliu77\}@stevens.edu, \\
\textsuperscript{[3]}\{zwang48, wjiang8\}@gmu.edu, 
\textsuperscript{[4]}yangjh1993@unm.edu, 
\textsuperscript{[5]}\{tong.geng, ang.li\}@pnnl.gov 
\vspace{-15pt}}
}


\maketitle

\begin{abstract}
Molecular similarity search has been widely used in drug discovery to identify structurally similar compounds from large molecular databases rapidly. With the increasing size of chemical libraries, there is growing interest in the efficient acceleration of large-scale similarity search. Existing works mainly focus on CPU and GPU to accelerate the computation of the Tanimoto coefficient in measuring the pairwise similarity between different molecular fingerprints. In this paper, we propose and optimize an FPGA-based accelerator design on exhaustive and approximate search algorithms. On exhaustive search using  BitBound \& folding, we analyze the similarity cutoff and folding level relationship with search speedup and accuracy, and propose a scalable on-the-fly query engine on FPGAs to reduce the resource utilization and pipeline interval. We achieve a 450 million compounds-per-second processing throughput for a single query engine. On approximate search using hierarchical navigable small world (HNSW), a popular algorithm with high recall and query speed. We propose an FPGA-based graph traversal engine to utilize a high throughput register array based priority queue and fine-grained distance calculation engine to increase the processing capability. Experimental results show that the proposed FPGA-based HNSW implementation has a 103385 query per second (QPS) on the Chembl database with 0.92 recall and achieves a 35$\times$ speedup than the existing CPU implementation on average. To the best of our knowledge, our FPGA-based implementation is the first attempt to accelerate molecular similarity search algorithms on FPGA and has the highest performance among existing approaches.

 
 
\end{abstract}


\section{Introduction}
In drug discovery, molecular similarity search \cite{maggiora2014molecular} has been widely used to  identify structurally similar compounds from large molecular databases rapidly. 
Unlike image or document search in multimedia data retrieval, chemists often compute the Tanimoto or Jaccard similarity coefficient to measure the pairwise similarity. 
The distance between 
a query and a known molecule database can be measured through the K nearest neighbor (KNN) similarity search.
In general, there are two types of search methods, exhaustive search, and approximate search. Exhaustive KNN search requires a linear scan through the entire database. 
Approximate methods reduce search complexity in exchange for 
search precision.
As the size of compound libraries increases, efficient similarity search techniques are highly desired.
Existing works mainly focus on using CPU and GPU
to accelerate the computation of the Tanimoto similarity coefficient. For instance, the BitBound algorithm~\cite{dalke2019chemfp} and the modulo-M BitBound algorithm~\cite{FPSim2} were developed for high-performance cheminformatics fingerprint computation on the CPU.  
GPUsimilarity \cite{gpusimilarity} provides the modulo-OR-compression (folding) algorithm implementation for the database compression on GPU. 

Due to extremely low latency, high energy efficiency, and flexible programmability for easy prototyping, FPGAs have received much attention as an alternative accelerating solution
for various data analytics applications \cite{jiang2020hardware, peng2021accelerating, qi2021accommodating, peng2021binary, jiang2019accuracy, peng2020selective}. 
However, there are few reports about the acceleration of molecular similarity search using FPGAs; in particular, the hardware-algorithm co-design exploration on FPGAs remains unknown. 

\begin{figure}[b]
\vspace{0pt}
\centering
\includegraphics[width=.48\textwidth]{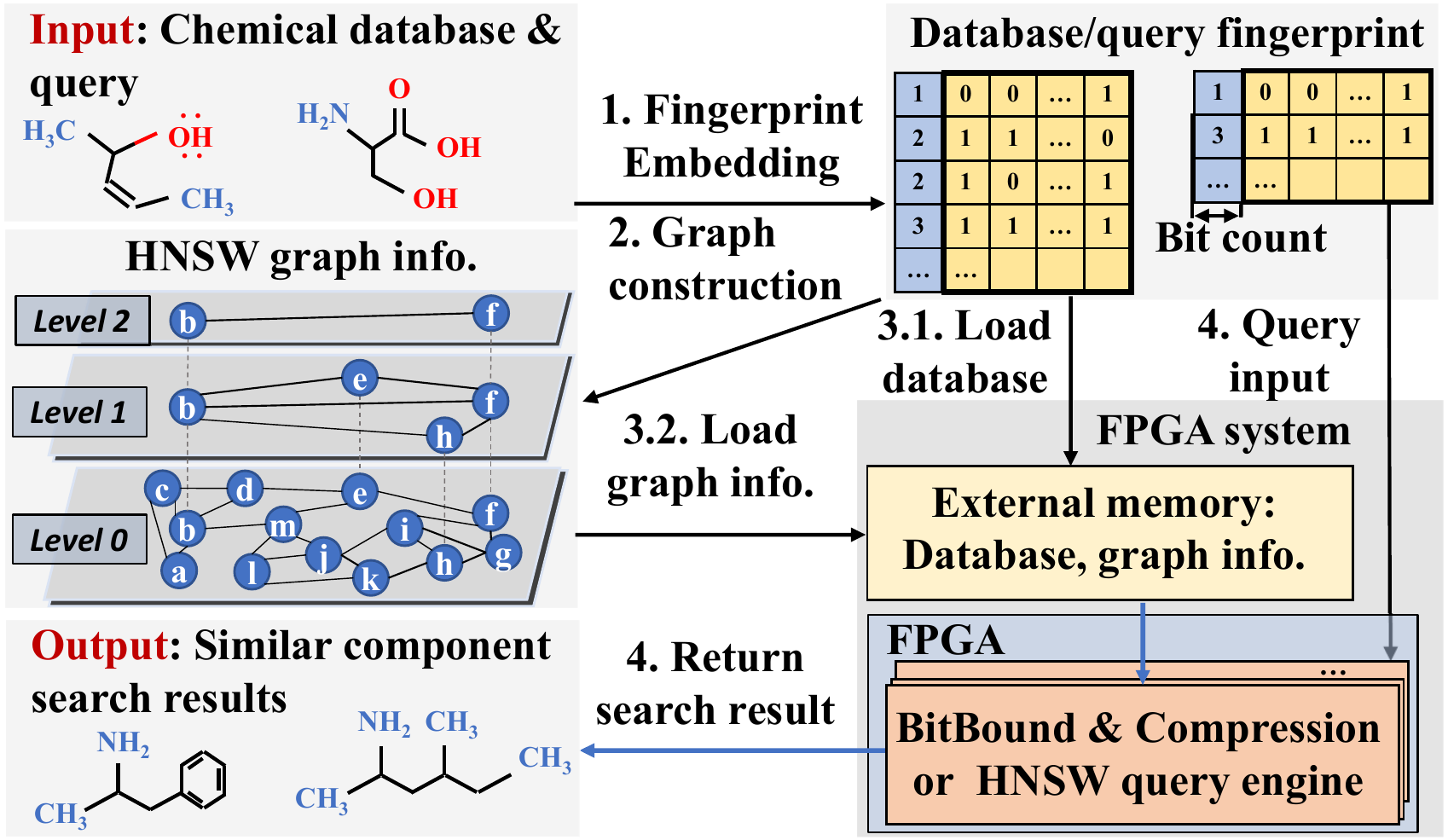}
\caption{Overview of FPGA-based Molecular Similarity Search. }
\label{fig:sys_stru}
\vspace{-5pt}
\end{figure}

In this work, to efficiently explore the FPGA design space, we propose an analytical design scheme.  We design and optimize the representative algorithm in exhaustive search (BitBound \& folding algorithm) and approximate search (HNSW), respectively. 
The overall of our design flow is shown in Fig.~\ref{fig:sys_stru}. Our proposed method outperforms existing approaches for two reasons. 
Firstly, work \cite{parravicini2021scaling} separates the distance calculation and the top-k sort operation, leading to high memory bandwidth consumption and low throughput. 
Second, the work \cite{zhang2018efficient} utilizes a parallel priority queue for the top-k search, but the scaling of k size will degrade the kernel frequency and lead to extensive resource usage. While we leverage the fine-grained data movement scheme to reduce the pipeline interval and uses top-k merge sort and top-k priority queue for the sorting. Our proposed design has a much higher kernel throughput and better scalability for the similarity search application.



Our contributions are summarized as follows: 

\begin{itemize}
    \item We propose and optimize FPGA-based accelerator designs on the exhaustive and approximate search for large-scale molecular similarity search.
    \item On exhaustive search using  BitBound \& folding, we 
    propose a scalable on-the-fly query engine on FPGAs to reduce the resource utilization and pipeline interval, and achieve a 450 million compounds/s processing throughput for a single query engine. For Chembl database, our accelerator has 1638 QPS throughput for brute-force search and 25403 QPS throughput for BitBound \& folding design with 0.97 recall. 
    \item We propose an FPGA-based graph traversal engine for HNSW approximate search. 
    We explore the relationship between the returned element size, adjacent list size, and processing throughput and develop a Pareto frontier. The proposed FPGA-based HNSW implementation has a 103385 QPS with 0.92 recall on Chembl database and achieves a 35$\times$ speedup than CPU on average. 
    
\end{itemize}

To the best of our knowledge, our FPGA-based implementation has the highest performance among existing approaches on molecular similarity search.

\section{Background and Related Works}


\subsection{Jaccard/Tanimoto similarity}
There are ECFP, Morgan circular \cite{morgan1965generation}, and MACCS structural keys \cite{durant2002reoptimization} fingerprints to measure molecular similarity.  
In this study, we adopt the 1024-bit Morgan binary fingerprint \cite{landrum2013rdkit}. 
For two given binary molecular fingerprints, $\vec{A}=(a_1,\ a_2\ldots a_{1024})$ and $\vec{B}=(b_1,\ b_2\ldots b_{1024})$ ($a_i, b_i \in \{0, 1\}$), their Jaccard/Tanimoto similarity\cite{real1996probabilistic} is the division of their intersection and union number of 1s (denoted using $\sum$): 
\begin{equation} 
S(\vec{A},\ \vec{B})\ = \frac{\sum(\vec{A}\cap \vec{B})}{\sum(\vec{A}\cup \vec{B})} 
\label{eq:TFC} 
\end{equation}

\subsection{Existing Works}
There are several existing frameworks for KNN search in image or document similarity search. Faiss \cite{JDH17, faiss} is a popular dense vector KNN search framework for HNSW implementations on CPU and GPU platforms. 
NMSLIB \cite{boytsov2013sisap,NMSLIB} is focused on providing support for distance measures in non-metric spaces. 
FALCONN \cite{andoni2015practical, FALCONN}
utilizes locality-sensitive hashing for approximate distance calculations.



For the molecule similarity search application, Chemfp \cite{dalke2019chemfp} and FPSim2\cite{FPSim2} implement the modulo-M BitBound algorithm. 
offering a tight bound on similarity score between a query and the database.
A benchmark \cite{zhu2020benchmark} is conducted for existing similar search algorithms implementation on Intel Xeon E5-2690 CPU platform. The benchmark on the Chembl database shows exhaustive search algorithms, such as brute-force and BitBound algorithms, have throughputs of 23 QPS and 46 QPS; the approximate search algorithms, such as HNSW and folding algorithms, have 950 QPS and 121 QPS throughputs while the recall is 0.9. 
GPUsimilarity \cite{gpusimilarity}  provides a GPU implementation of brute-force chemical similarity search  and has a throughput of 570 QPS on chembl database. 

\section{Indexing Algorithms}


In this section, we will first discuss the algorithm selection process and criteria for molecular similarity search on the FPGA. 
Secondly, we will provide a detailed analysis of the presented Bitbound \& Folding and HNSW algorithms. 


\subsection{Algorithm selection}
The molecule similarity search normally requires algorithms to have a high accuracy rate. 
BitBound \cite{swamidass2007bounds} algorithm provides an $O(n\textsuperscript{0.6})$ speedup by reducing the search space and is an exhaustive similarity search algorithm. 
For the large fingerprint database, 
a compression scheme is needed.
The modulo-OR-compression (folding) algorithm is the most commonly used algorithm in molecule similarity search \cite{nasr2010hashing, FPSim2}. Those algorithms are combined as BitBound \& folding algorithm. 


Besides the above algorithm, {the HNSW index algorithm performs excellently in recall and query speed \cite{zhu2020benchmark, malkov2018efficient} and is used for the approximate KNN search. 
The HNSW constructs a hierarchical graph index for approximate nearest neighbor search with poly-logarithmic construction complexity and logarithmic search complexity. 
Since the HNSW \cite{malkov2018efficient} algorithm was introduced in 2016, it has been one of the fastest high precision approximate nearest neighbors techniques.
}
The high recall of the HNSW algorithm is because it constructs a relative neighborhood graph, which has a heuristic algorithm for neighbor selection. The heuristic keeps a long-range link to help prevent a search from getting stuck in
local optima. 

\subsection{Analysis of BitBound and Folding Algorithms}
\label{section:BitBound}


This section uses the Chembl 27.1 database \cite{mendez2019chembl} with 1.9 million molecules for the analysis. The RDKit toolbox is used for fingerprint generation. 
The accuracy (recall) 
is defined as the Top-K search matching rate between the proposed and brute-force algorithms.

The BitBound algorithm record the number of bits for each database fingerprint and query fingerprint and use the similarity cutoff ($S_c$) and bound inequality to reduce the search space. 
For query compound fingerprint $\vec{A_i}$, only the database query $B_j$ which satisfies the following inequation is searched:





\begin{equation}
Cnt(\vec{A_i}) \cdot S_c \leqslant Cnt(\vec{B_i}) \leqslant Cnt(\vec{A_i})/S_c
\label{eq:BitBound}
\end{equation}

The bit count probability distribution of the Chembl database is modeled as Gaussian distribution $ x \sim \mathcal{N}(\mu,\,\sigma^{2}) $: 


\begin{equation}
f_g(x)=\frac{1}{\sqrt{2\pi\sigma^2}}e^\frac{-{(x-\mu)}^2}{2\sigma^2}
\label{eq:Gua1}
\end{equation}

\begin{figure}[h!]
    \centering
    \centering
\begin{multicols}{2}
\vspace{-10pt}
\subfloat  [\label{fig:Prob_dis} Probability distribution. ]  {\includegraphics[width=1.03\linewidth]{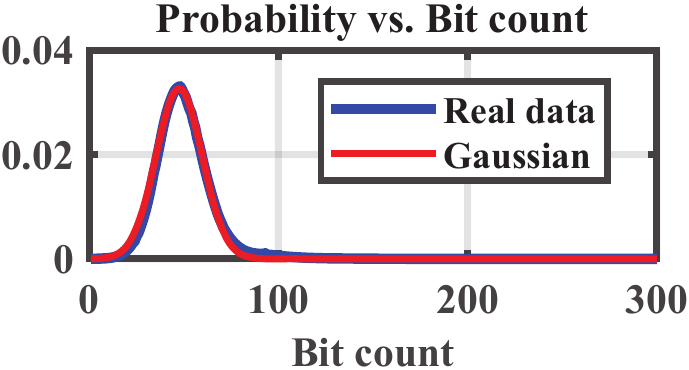}\par }
\subfloat  [\label{fig:Prune_SC03} Pruned search space ($S_c$ = 0.3). ]  {\includegraphics[width=0.97\linewidth]{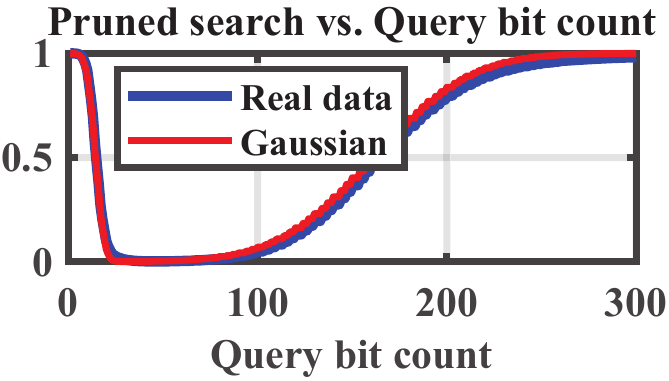}\par }

\end{multicols}
\vspace{-25pt}
\begin{multicols}{2}
\subfloat [\label{fig:Prune_SC08} Pruned search space ($S_c$ = 0.8). ]   {\includegraphics[width=1.01\linewidth]{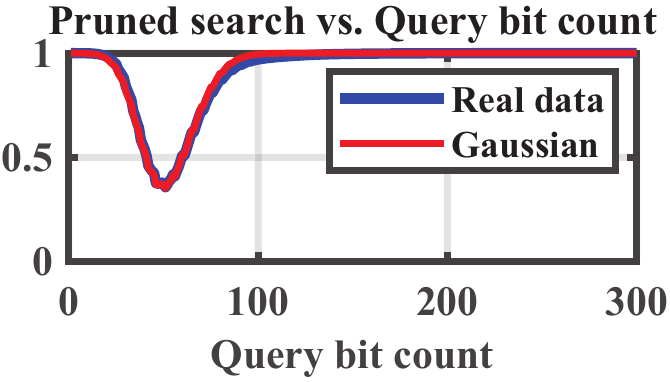}\par }
\subfloat  [\label{fig:Speedup_sc} Average speedup. ] {\includegraphics[width=0.99\linewidth]{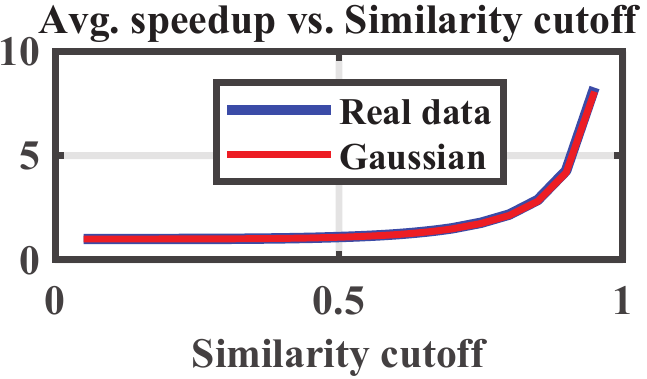}\par }
\end{multicols}
    \vspace{-5pt}
    \caption{Modeling of the BitBound algorithm. }
    \label{fig:bound_space}
\end{figure}

Variables $x$, $\mu$, and $\sigma$ represent the fingerprint bit count, the average bit count, and the standard deviation. 
The pruned search space for 
similarity cutoff $S_c = 0.3$ and $S_c = 0.8$ comparisons are given in Fig. \ref{fig:Prune_SC03} and Fig. \ref{fig:Prune_SC08}. 
The final speedup vs. similarity cutoff relationship is given in Fig. \ref{fig:Speedup_sc}. The actual speedup increase with the increase of the similarity cutoff. 

\begin{figure}[t]
\centering
\includegraphics[width=.4\textwidth]{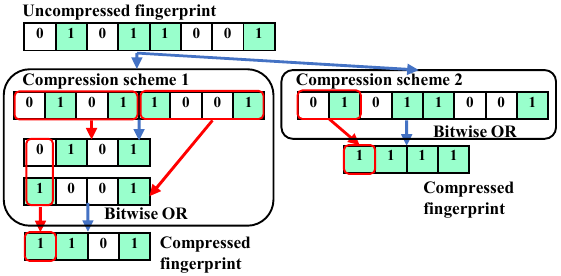}
\caption{2 compression schemes example (L = 8, m = 2). }
\label{fig:compression_scheme}
\end{figure}




There are two types of folding schemes. 
Assuming the fingerprint length is L, and the folding level is m. 
The first scheme conducts the bitwise OR between L/m section, and the second scheme conducts the bitwise OR between every nearby m bits. An example is shown in Fig. \ref{fig:compression_scheme}. 
A 2-stage folding algorithm \cite{gpusimilarity} is used to retain the accuracy. 
The first search is conducted on the compressed database, and the top $k_{r1}$ results are returned. The second search is conducted on returned indices from the first round and the uncompressed database, and $k$ elements are returned. The relationship between 
$k_{r1}$, 
$k$, and $m$ can be expressed as $k_{r1} = k \cdot mlog_{2}(2m)$. 
The final accuracy test results of Top-20 search can be found in Table \ref{tab:compression_tb}, and compression scheme 1 has higher accuracy than compression scheme 2. 
Thus the compression scheme 1 is used. 


\begin{table}[h!]
    \caption{Accuracy vs. folding level (m)}

    \centering
    \begin{tabular}{cccc}
        \hline
                m             & Folding 1   & Folding 2 & $m·log_2(2m)$ \\
                             & accuracy (\%)  &  accuracy (\%) &  \\
         \hline
     1 & 100 & 100 & 1   \\
     \hline
     2 & 99.3 & 91.5 & 4   \\
        \hline
     4 & 99.1 & 92.1 & 12   \\
      \hline
     8 & 97.3 & 89.2 & 32   \\
        \hline
     16 & 84.4 & 76.2 & 80   \\
      \hline
     32 & 31.7 & 31.1 & 192   \\
      
        \hline
    \end{tabular}
    \vspace{-5pt}
        \label{tab:compression_tb}
\end{table}


\subsection{Analysis of HNSW Algorithm}


The HNSW uses two tuning parameters to control the quality of the constructed index: returned elements count $E_f$ adjacency list size and $M$. 
A larger value $E_f$ may help the search escape local optima by requiring the search to select a compulsory $E_f$ element into a candidate list. 
A greater value of $M$ will give higher recall but will be required up to 2M distance calculations per iteration in the graph traversal, which will affect the overall throughput of search. The Hnswlib \cite{HNSWLIB} implementation also provides a parallel construction algorithm that allows for multiple elements to be inserted into the graph simultaneously. Due to memory bandwidth limitations and the need for parallel guards, the parallel construction algorithm achieves logarithmic scaling. KNN search can be performed on the constructed graph index. Search can use the $E_f$ parameter similarly to construction to control the batch quality at the cost of longer search times. However, a high-quality graph index, a graph built with higher $E_f$, may provide sufficiently high recall in search with $E_{fsearch} <  E_{fconstruction}$.

\section{Optimizing Indexing Algorithms on FPGA}

This section will demonstrate the proposed FPGA designs to optimize the indexing algorithms for molecule similarity search.
We target two categories of indexing algorithms and select one typical algorithm for each category: (1) the combination of BitBound and Folding Algorithm (denoted as BitBound \& folding) \cite{swamidass2007bounds}, which can support the exhaustive search; and (2) Hierarchical Navigable Small World (HNSW) \cite{malkov2018efficient}, which is a representative of approximate search.
Kindly note that the proposed FPGA design can be easily extended to other algorithms in the same category.

\subsection{FPGA design to support exhaustive search}
\begin{figure}[t]
\centering
\includegraphics[width=.45\textwidth]{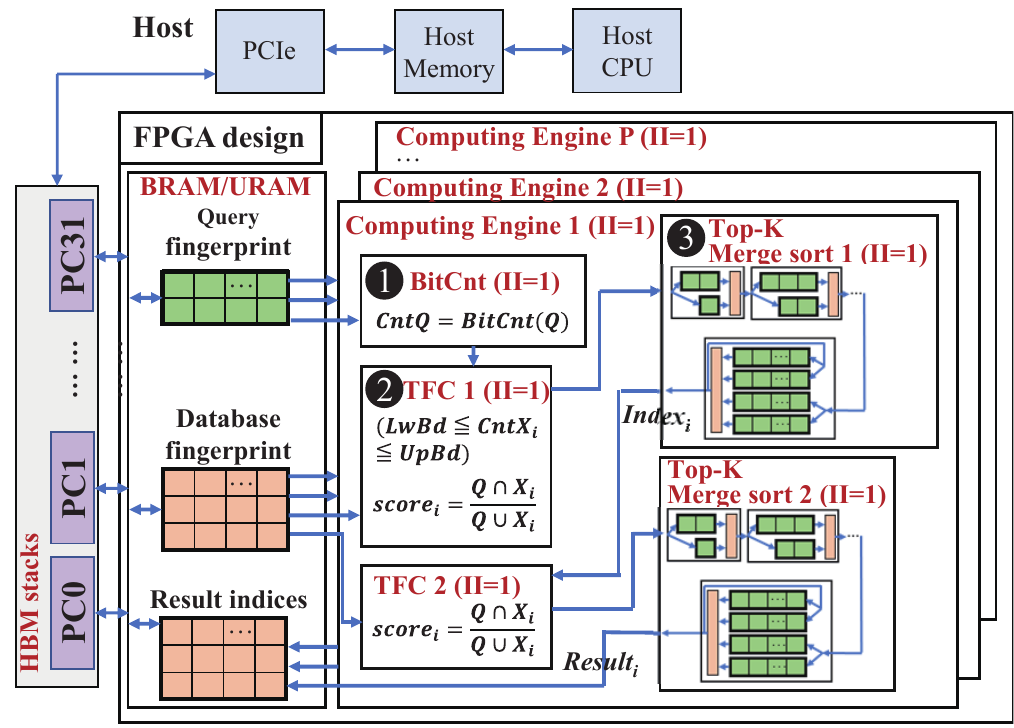}
\caption{FPGA design of indexing algorithm for exhaustive similarity search based on the BitBound \& folding algorithm.}
\label{fig:Kernal_design_BC}
\end{figure}

Figure \ref{fig:Kernal_design_BC} demonstrates the overview of the FPGA design for the exhaustive similarity search algorithm.
The full design can be divided into two sub-systems: (1) the communication sub-system and (2) the computation sub-system.
Unlike the existing design applying the sequential process between two sub-systems \cite{danopoulos2019approximate}, we propose the design to enable the computation and communication to be conducted in a pipelined fashion for the similarity search, which is called ``on-the-fly'' design in this paper.
We adopt the FPGA boards with high bandwidth memory (HBM) to resolve the communication bottleneck in this work. 
For the computation sub-system, we develop three components to support the indexing algorithm: \ding{182} Bit Count module, denoted as BitCnt, \ding{183} Tanimoto Factor Calculation module, denoted as TFC, and \ding{184} Top-K module.

\noindent\ding{182} \textbf{BitCnt}: The BitCnt kernel counts the number of bits for the input binary fingerprint. The resource utilization of the BitCnt kernel will scales linearly with the binary fingerprint length. 

\noindent\ding{183} \textbf{TFC}: The Tanimoto factor calculation (TFC) module calculates the similarity score between query and database. In order to reduce the computation and storage overhead without loss of accuracy, the Tanimoto score information is stored as 12 bits fixed-point. According to Eq. \ref{eq:TFC}, the Tanimoto factor calculation includes 2 bit count accumulation kernels and 1 fixed-point division operation. 


\noindent\ding{184} \textbf{Top-K merge}: The top-k merge is built with FIFO and comparator.
Unlike the existing top-K module, which is typically based on the parallel priority queue \cite{zhang2018efficient}, our proposed top-k is based on the merge sort structure.
As such, we can fully utilize HBM for fast memory access. In addition, we can achieve higher throughput and low resource utilization with the scaling of the k size (observation 2).

The Top-K sorter utilizes $log_2K + 1$ comparators and $log_2K + 2K$ FIFO capacity. The small size FIFO can be built upon the register, and the large size FIFO can be built BRAM block. The pipeline interval is tuned as 1 for high throughput, and the entire implementation has a latency of $N + log_2K$ with $N$ input sequence size. 
The resource utilization roughly scales in a $O(log(k))$ size. For all different k values, the critical path has been tuned to achieve a 450 MHz clock frequency. 



\noindent\textbf{Put it all together}: The computing Engine in Fig. \ref{fig:Kernal_design_BC} shows how these components work together.
Specifically, the fingerprint fetch, TFC kernel, and top-k merge sort kernel are cascaded to achieve an on-the-fly data movement structure. With the fine-grained TFC and top-k merge sort kernel design, the overall implementation achieves a pipeline interval as 1 for the cascaded structure. In order to achieve a high query speed, the HBM is utilized for the data movement. 
The top-k search kernel can run at a high frequency (450MHz) and consumes 57.6 GB/s memory bandwidth in our implementation.

\subsection{FPGA design to approximate search}

\begin{figure}[t]
\centering
\includegraphics[width=.45\textwidth]{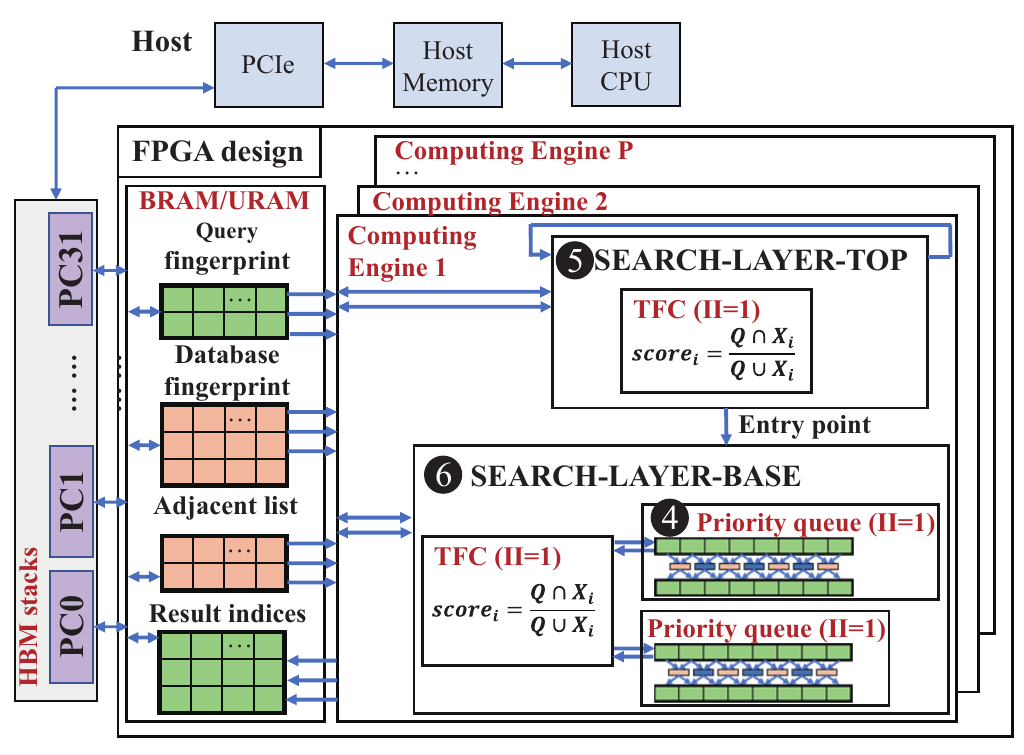}
\caption{FPGA design of indexing algorithm for the approximate similarity search based on the HNSW algorithm.}
\label{fig:Kernal_design_HNSW}
\end{figure}

We adopt a similar FPGA architecture in the design of the accelerator for the approximate molecule similarity search.
Figure \ref{fig:Kernal_design_HNSW} demonstrates our proposed design.
In such a design, we adopt the proposed \ding{183} TFC module in Figure \ref{fig:Kernal_design_BC} and proposed \ding{185} Priority Queue (PQ) module, which will be the basic component to support the computing engine.
On top of this, we further propose two modules to support the HNSW algorithm: \ding{186} Search-Layer-Top module and \ding{187} Search-Layer-Base module.

\noindent\ding{185} \textbf{PQ}: Instead of merge sort, the priority queue is another method to support the Top-k operation.
Our design philosophy is to achieve a high throughput design without frequency degradation.
In consequence, we employ the register array structure to implement the priority queue.
The compare and swap operation is done between even and old entries for each clock cycle, and the kernel operation achieves a pipeline interval as 1 for both enqueue and dequeue operation. Thus, the throughput of the hardware priority queue is as high as clock frequency. The number of comparators scales linearly with the size of the priority queue. Thus the register array design is not favored when the priority queue size is large. 


Assuming the entry size of the top-k priority queue is a 12 bit fixed-point number, the resource utilization with a different scale of k can be obtained. The resource utilization of the top-k priority queue is bounded most heavily by the LUT. The FF and LUT utilization scales linearly with the size of k. 
 
\noindent\ding{186} \textbf{Search-Layer-Top module}: HNSW graph construction is based on the randomly shuffled database with index, and then the graph adjacent lists information will be used for the HNSW search step.
The constructed HNSW comprises multiple layers, and each layer will have its own set of adjacency lists. During the HNSW graph search, the graph traverse Algorithm \ref{alg:alg1} will traversal all the layers except the base layer.

\begin{algorithm}[h!] 
\caption{ SEARCH-LAYER-TOP($q_r$, $e_p$, $l_i$)}
\label{alg:alg1} 
\begin{algorithmic}[1] 
\REQUIRE 
query fingerprint $q_r$, entry point fingerprint $e_p$, layer number $l_i$\\
\ENSURE 
Closest neighbor to query $top$\\
\STATE 	$C = ep$  $\verb|//|$ Candidate element
\STATE 	$C_n = q$  $\verb|//|$ Next candidate element
 \WHILE{$C \ != \ C_n$}
 \STATE $C = C_n$
 \FOR{each $e$ in $neighbourhood(C)$ at layer $l_i$}
\STATE $\verb|//|$ $distance$ function utilizes TFC kernel
 \IF{$distance(C, q) > distance(e, q)$} 
  \STATE $C_n = e$
 \ENDIF
 \ENDFOR
 \ENDWHILE
 \STATE $top = C$
 \RETURN $top$
\end{algorithmic}
\end{algorithm}

 
 
\noindent\ding{187} \textbf{Search-Layer-Base module}: The graph traversal Algorithm \ref{alg:alg2} will be conducted on the base layer, and top Ef results will be returned. The final results will be obtained from the top-k result of the Ef returned results using a simple top-k search. 


\begin{algorithm}[h!] 
\caption{ SEARCH-LAYER-BASE($q_r$, $e_p$, $ef$)}
\label{alg:alg2} 
\begin{algorithmic}[1] 
\REQUIRE 
query fingerprint $q_r$, entry point fingerprint $e_p$, number of returned elements $e_f$\\
\ENSURE 
$e_f$ nearest neighbors to query\\
\STATE 	$v \leftarrow e_p$  $\verb|//|$ $v$: visited elements set
\STATE 	$C \leftarrow e_p$  $\verb|//|$ $C$: candidates set
\STATE 	$M \leftarrow e_p$  $\verb|//|$ $M$: returned elements set
\STATE $\verb|//|$ $M$ and $C$ are built upon hardware priority queue
\WHILE{$size(C) > 0$}
\STATE $top \leftarrow$ get and pop closest element between $C$ and $q_r$
\STATE $fur \leftarrow$ get furthest element between $M$ and $q_r$
\IF{$distance(top,q_r)>distance(fur,q_r)$}
\STATE $\verb|//|$ $distance$ function utilizes TFC kernel
\STATE break $\verb|//|$ no further traversal is required
\ENDIF
\FOR{each $e \in neighbourhood(top)$}
\IF{$e \notin v$}
\STATE $v \leftarrow v \cup e$
\STATE $f \leftarrow$ get furthest element between $M$ and $q_r$
\IF{$distance(e, q_e) < distance (f, q_e)$ or \\ $size(M) < e_f$}
\STATE $\verb|//|$ $distance$ function utilizes TFC kernel
\STATE $C \leftarrow C \cup e$
\STATE $M \leftarrow M \cup e$
\IF{$size(W) > e_f$}
\STATE pop furthest element between $M$ and $q_r$
\ENDIF
\ENDIF
\ENDIF
\ENDFOR
\ENDWHILE
 \RETURN $M$
\end{algorithmic}
\end{algorithm}

\noindent\textbf{Put it all together:} The FPGA hardware design of the HNSW index algorithm can be found in Fig. \ref{fig:Kernal_design_HNSW}. Algorithm \ref{alg:alg2} \cite{malkov2018efficient} utilizes 2 register arrays based priority queue, and both of the priority queues are sized as $e_f$. Both algorithm \ref{alg:alg1} and algorithm \ref{alg:alg2} utilize only a single TFC kernel.

\section{Experimental and Design Exploration}


In this section, the experiment setup is firstly introduced. Then, we report the results on the design exploration and cross-platform comparison. Results will show that our optimized FPGA accelerator can achieve a 35$\times$ speedup over the existing acceleration on the general-purpose computing platform.

\subsection{Experimental Setup}

The Chembl 27.1 database \cite{mendez2019chembl} with 1.9 million molecules is used for the similarity search. The RDkit toolbox \cite{landrum2013rdkit} is used 
for binary fingerprint generation. 

The FPGA designs are implemented in Vivado HLS v2020.1. The FPGA hardware platform is the Alveo U280 board. Alveo U280 board is equipped with 8 GB HBM, and its maximum memory bandwidth is 460 GB/s. 
The memory bandwidth for linear memory access has been limited to under 410 GB/s to provide suitable overhead. Alveo U280 on-chip resource includes 960 URAM blocks, 4032 BRAM blocks, 9024 DSP48E, 2.6M FF, and 1.3M LUT. With the abundant HBM memory bandwidth and on-chip resources, more kernels can improve overall throughput. Thus, design space exploration is needed to reveal the trade-off between different design parameters.

\subsection{FPGA Design Exploration and Cross-work Comparison}

\subsubsection{FPGA Design Exploration}
The brute force similarity search kernel consumes around 0.4\% of the total LUT resource. The memory bandwidth requirement for a single brute force search kernel is 57.6 GB/s, and
7 kernels can be used to accelerate the single query request. The memory bandwidth bounds the maximum throughput for the brute-force similarity search. Each kernel runs at 450 MHz frequency, 
and the entire FPGA-based query engine achieves a 1638 QPS throughput. 

\begin{figure}[h!]\vspace{-15pt}
    \centering
    \centering
\begin{multicols}{2}
\subfloat  [\label{fig:Bit_fol_res} Resource utilization. ]  {\includegraphics[width=1\linewidth]{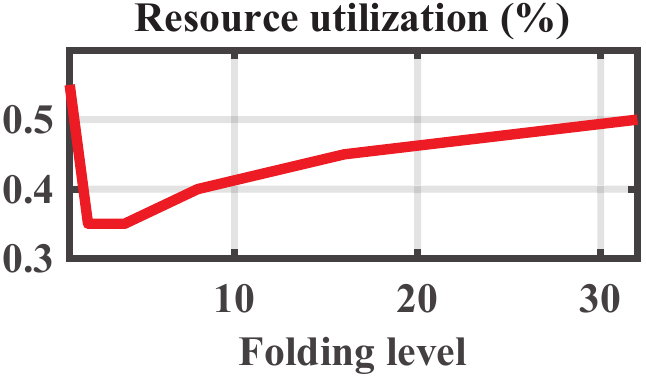}\par }
\subfloat  [\label{fig:Bit_fol_mem} Memory bandwidth requirement. ]  {\includegraphics[width=1\linewidth]{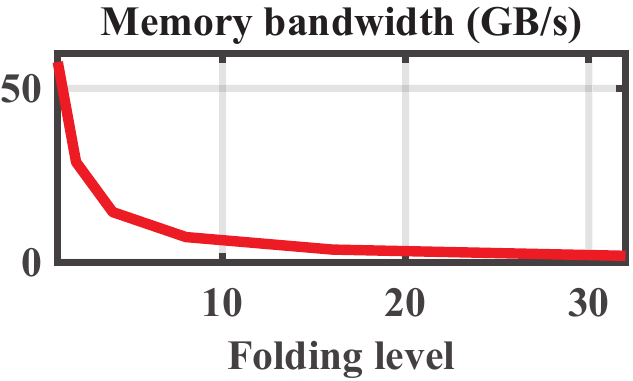}\par }

\end{multicols}

    \vspace{-10pt}
    \caption{Resource utilization and memory bandwidth requirement for the BitBound \& folding algorithm. }
    \label{fig:ABC}
\end{figure}

For the BitBound \& folding algorithm, the resource utilization  of a single kernel 
can be extracted from the Vivado HLS toolbox. The resource utilization (Bounded by LUT \& BRAM) vs. folding level 
can be found in Fig. \ref{fig:Bit_fol_res}. 
With the increase of the folding level, resource utilization firstly decreases and then increases. The later increase in resource utilization is due to the growing size of the merge sort circuit. 
The memory consumption can be found in Fig. \ref{fig:Bit_fol_mem}. 
The memory consumption decreases with the increase of the folding level. 

\begin{figure}[h!]
\centering
\includegraphics[width=.35\textwidth]{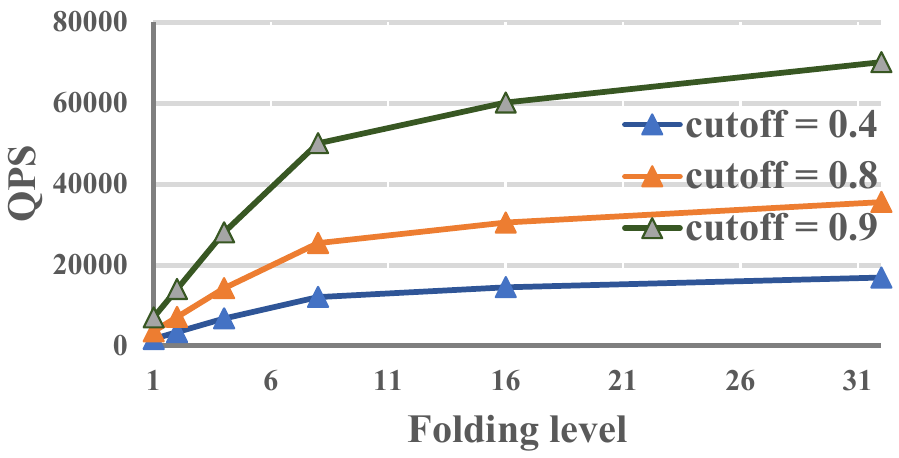}
\caption{QPS for BitBound \& folding algorithm on FPGA. }
\label{fig:QPS_comp_lel}
\end{figure}

The search space reduction of the BitBound algorithm can be found in Fig. \ref{fig:bound_space}. 
Thus, the BitBound \& folding algorithm throughput on the FPGA platform can be found in Fig. \ref{fig:QPS_comp_lel}. With the increase of the folding level, the query speed increases. 




\begin{figure}[h!]
\centering
\includegraphics[width=.35\textwidth]{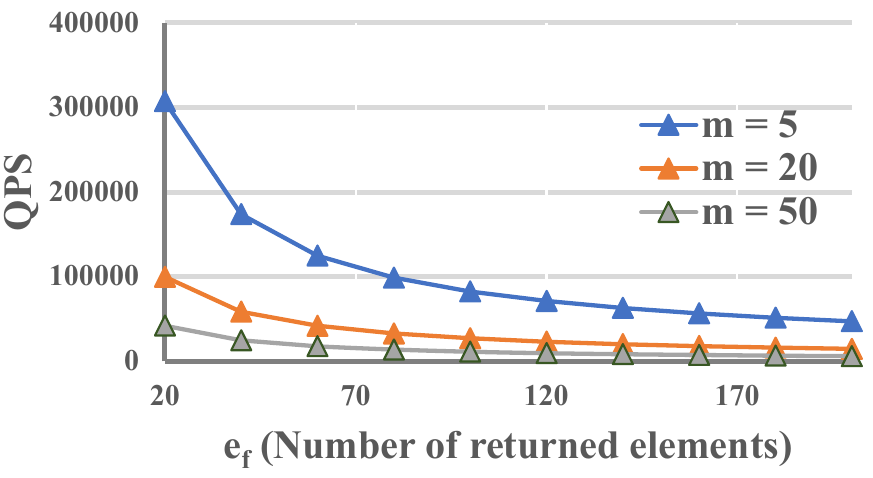}
\caption{QPS for HNSW algorithm.  }
\label{fig:QPS_HNSW_m_ef}
\end{figure}

The LUT resources majorly dominate the resource usage of the HNSW query engine. With the increase of the number of returned elements $e_f$, the LUT resource usage increases. The input parameter $m$ for HNSW is the maximum size of the adjacent list within the upper layers of the graph. The base layer of the graph provides every element up to $2M$ adjacency list elements, of which a minimum of $M$ adjacency list items are filled after a vertex is inserted. Both of the parameters above, $e_f$ and $m$, will affect the throughput of the query engine. A design exploration study is conducted based on different $e_f$ and $m$ parameters. The QPS vs. $m$ and $e_f$ can be found in Fig. \ref{fig:QPS_HNSW_m_ef}. The design exploration reveals that the query speed increases with the decrease of both $m$ and $e_f$ design parameters. 

\subsubsection{Algorithms Performance on FPGA Platform}

\begin{figure}[h!]
\centering
\includegraphics[width=.35\textwidth]{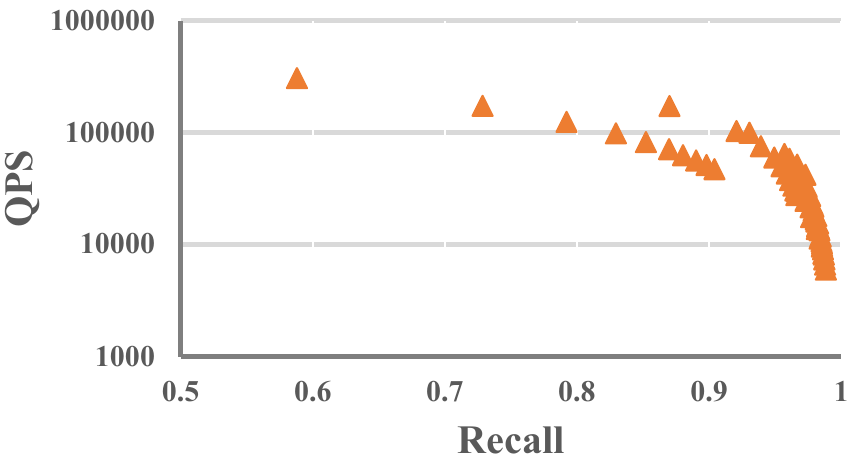}
\caption{Design exploration for HNSW algorithm. }
\label{fig:QPS_HNSW_explo}
\end{figure}

To reveal the true performance of the HNSW algorithm on the FPGA platform, a grid search for $e_f$ and $m$ is conducted. The $m$ is set to be 5, 10, 20 ..., 50, and $e_f$ is set to be 20, 40, 60 ..., 200. For all of the hyperparameter combinations, the QPS and recall are recorded. The design space exploration is conducted, and the QPS vs. recall 
for different hyper-parameters 
is shown in Fig. \ref{fig:QPS_HNSW_explo}.

\begin{figure}[h!]
\centering
\includegraphics[width=.35\textwidth]{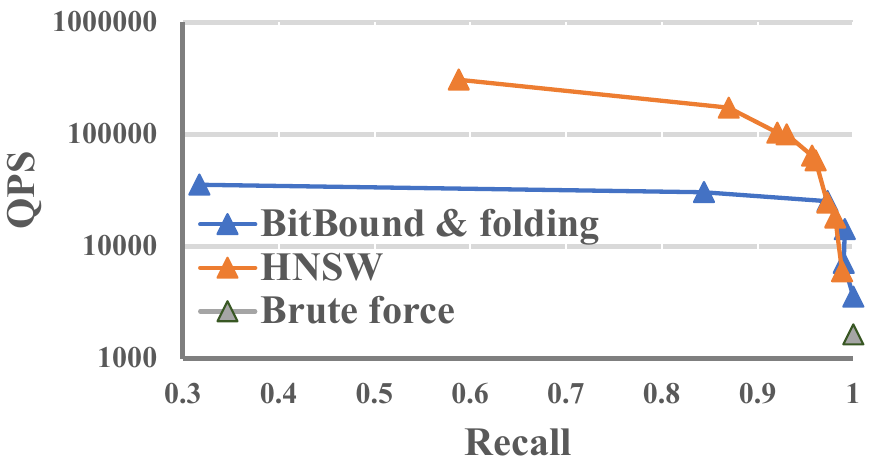}
\caption{Pareto frontiers of different algorithms on FPGA. }
\label{fig:QPS_pareto}
\vspace{-5pt}
\end{figure}

As shown in Fig. \ref{fig:QPS_pareto}, the Pareto frontier is generated based on the design exploration of the above algorithms. The similarity cutoff of BitBound \& folding algorithms is set at 0.8. While the recall is high, the QPS for BitBound \& folding algorithm is better. With a lower recall requirement, the HNSW algorithm has a much better performance.

\subsection{Cross-platform Comparison}

The CPU implementation in this section is based on  Intel Xeon Gold 6244 CPU platform, and GPU implementation in this section is based on two NVIDIA Tesla V100. 

The CPU implementation of brute force, BitBound \& folding, and HNSW algorithms can be found in \cite{zhu2020benchmark}.
We utilize  the code and method to run on our platform.  
The same as the FPGA implementation, the design exploration is conducted on the CPU platform. The GPU implementation of the brute-force similarity search method \cite{gpusimilarity} is also included. The fingerprint database is loaded into the main memory for both implementations, and the search will be conducted. 

\begin{figure}[h!]
\centering
\includegraphics[width=.35\textwidth]{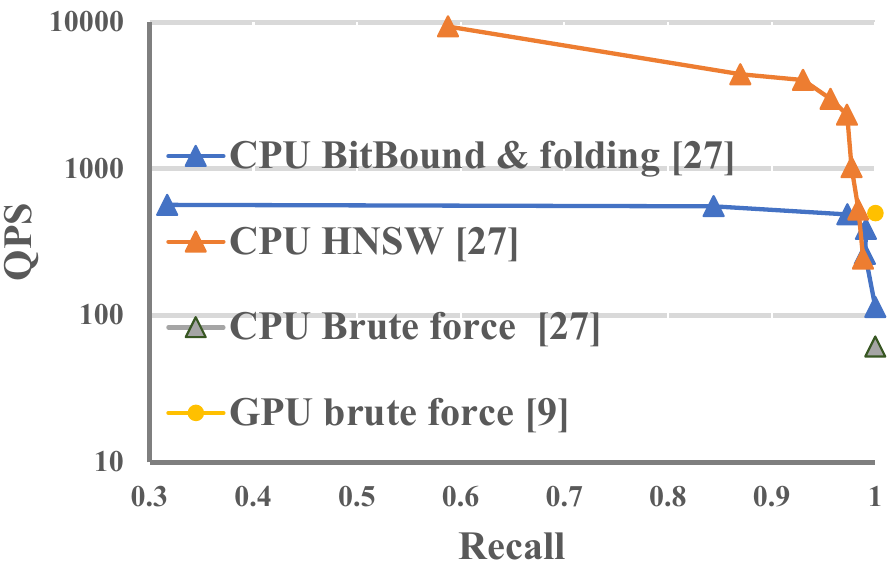}
\caption{Pareto frontier for different CPU/GPU platforms. }
\label{fig:QPS_pareto_CPUGPU}
\end{figure}

The Pareto frontier curve obtained from the design space exploration is shown 
in Fig. \ref{fig:QPS_pareto_CPUGPU}. The GPU platform outperforms the CPU platform on the brute force search algorithm. However, the CPU platform achieves similar performance to the GPU platform without much recall degradation when the BitBound \& compression algorithm is implemented. When the recall requirement is lower, the HNSW algorithm on the CPU platform has a much higher throughput than the BitBound \& folding compression algorithm. 

Comparing the FPGA platform implementation (Fig. \ref{fig:QPS_pareto}) and CPU/GPU platform implementation (Fig. \ref{fig:QPS_pareto_CPUGPU}), we find out the FPGA implementation achieves more than 25× speedup on brute force algorithm over CPU platform and more than 3× speedup than the GPU platform. For the BitBound \& folding compression index algorithm, the FPGA implementation achieves an average 30× speedup than the CPU platform. For the HNSW index algorithm, the FPGA implementation achieves an average 35× speedup than the CPU platform. 






\section{Conclusion}


The paper proposes the design and exploration for the different molecule similarity search algorithms on FPGAs. We design and optimize the two representative works in exhaustive search, BitBound \& Folding algorithms and approximate search, HNSW, respectively. More specifically, on BitBound \& folding, we analyze the similarity cutoff and folding level relationship with search speedup/accuracy and propose a scalable on-the-fly query engine on FPGAs to reduce the resource utilization and pipeline interval. 
On approximate search using HNSW, we propose an FPGA-based graph traversal engine 
to utilize high throughput register array based priority queue and fine-grained distance calculation engine to increase the processing capability. We further explore the relationship between the returned element size,   adjacent list size,   and the processing throughput and develop a Pareto frontier based on the design space exploration. Experimental  results  show  that  
our FPGA-based implementation, to the best of our knowledge, is the first attempt to accelerate molecular similarity search on FPGA and achieves the highest performance among existing works.



{
\bibliographystyle{IEEEtran}
\bibliography{bibligraphy}

\begin{thebibliography}{10}
\providecommand{\url}[1]{#1}
\csname url@samestyle\endcsname
\providecommand{\newblock}{\relax}
\providecommand{\bibinfo}[2]{#2}
\providecommand{\BIBentrySTDinterwordspacing}{\spaceskip=0pt\relax}
\providecommand{\BIBentryALTinterwordstretchfactor}{4}
\providecommand{\BIBentryALTinterwordspacing}{\spaceskip=\fontdimen2\font plus
\BIBentryALTinterwordstretchfactor\fontdimen3\font minus
  \fontdimen4\font\relax}
\providecommand{\BIBforeignlanguage}[2]{{%
\expandafter\ifx\csname l@#1\endcsname\relax
\typeout{** WARNING: IEEEtran.bst: No hyphenation pattern has been}%
\typeout{** loaded for the language `#1'. Using the pattern for}%
\typeout{** the default language instead.}%
\else
\language=\csname l@#1\endcsname
\fi
#2}}
\providecommand{\BIBdecl}{\relax}
\BIBdecl

\bibitem{maggiora2014molecular}
G.~Maggiora, M.~Vogt, D.~Stumpfe, and J.~Bajorath, ``Molecular similarity in
  medicinal chemistry: miniperspective,'' \emph{Journal of medicinal
  chemistry}, vol.~57, no.~8, pp. 3186--3204, 2014.

\bibitem{dalke2019chemfp}
A.~Dalke, ``The chemfp project,'' \emph{Journal of Cheminformatics}, vol.~11,
  no.~1, pp. 1--21, 2019.

\bibitem{FPSim2}
E.~Félix \emph{et~al.}, ``Fpsim2: Simple package for fast molecular similarity
  searches,'' \url{https://github.com/chembl/FPSim2}, 2020.

\bibitem{gpusimilarity}
P.~Lorton, ``Brute-force gpu implementation of chemical fingerprint similarity
  searching.'' \url{https://github.com/schrodinger/gpusimilarity}, 2019.

\bibitem{jiang2020hardware}
W.~Jiang, L.~Yang, E.~H.-M. Sha, Q.~Zhuge, S.~Gu, S.~Dasgupta, Y.~Shi, and
  J.~Hu, ``Hardware/software co-exploration of neural architectures,''
  \emph{IEEE Transactions on Computer-Aided Design of Integrated Circuits and
  Systems}, vol.~39, no.~12, pp. 4805--4815, 2020.

\bibitem{peng2021accelerating}
H.~Peng, S.~Huang, T.~Geng, A.~Li, W.~Jiang, H.~Liu, S.~Wang, and C.~Ding,
  ``Accelerating transformer-based deep learning models on fpgas using column
  balanced block pruning,'' in \emph{2021 22nd International Symposium on
  Quality Electronic Design (ISQED)}.\hskip 1em plus 0.5em minus 0.4em\relax
  IEEE, 2021, pp. 142--148.

\bibitem{qi2021accommodating}
P.~Qi, Y.~Song, H.~Peng, S.~Huang, Q.~Zhuge, and E.~H.-M. Sha, ``Accommodating
  transformer onto fpga: Coupling the balanced model compression and
  fpga-implementation optimization,'' in \emph{Proceedings of the 2021 on Great
  Lakes Symposium on VLSI}, 2021, pp. 163--168.

\bibitem{peng2021binary}
H.~Peng, S.~Zhou, S.~Weitze, J.~Li, S.~Islam, T.~Geng, A.~Li, W.~Zhang,
  M.~Song, M.~Xie \emph{et~al.}, ``Binary complex neural network acceleration
  on fpga,'' in \emph{2021 IEEE 32nd International Conference on
  Application-specific Systems, Architectures and Processors (ASAP)}.\hskip 1em
  plus 0.5em minus 0.4em\relax IEEE, 2021, pp. 85--92.

\bibitem{jiang2019accuracy}
W.~Jiang, X.~Zhang, E.~H.-M. Sha, L.~Yang, Q.~Zhuge, Y.~Shi, and J.~Hu,
  ``Accuracy vs. efficiency: Achieving both through fpga-implementation aware
  neural architecture search,'' in \emph{Proceedings of the 56th Annual Design
  Automation Conference 2019}, 2019, pp. 1--6.

\bibitem{peng2020selective}
H.~Peng, B.~Narayanasamy, A.~I. Emon, Z.~Yuan, R.~Zhang, and F.~Luo,
  ``Selective digital active emi filtering using resonant controller,'' in
  \emph{2020 IEEE International Symposium on Electromagnetic Compatibility \&
  Signal/Power Integrity (EMCSI)}.\hskip 1em plus 0.5em minus 0.4em\relax IEEE,
  2020, pp. 632--639.

\bibitem{parravicini2021scaling}
A.~Parravicini, L.~G. Cellamare, M.~Siracusa, and M.~D. Santambrogio, ``Scaling
  up hbm efficiency of top-k spmv for approximate embedding similarity on
  fpgas,'' \emph{arXiv preprint arXiv:2103.04808}, 2021.

\bibitem{zhang2018efficient}
J.~Zhang, S.~Khoram, and J.~Li, ``Efficient large-scale approximate nearest
  neighbor search on opencl fpga,'' in \emph{Proceedings of the IEEE Conference
  on Computer Vision and Pattern Recognition}, 2018, pp. 4924--4932.

\bibitem{morgan1965generation}
H.~L. Morgan, ``The generation of a unique machine description for chemical
  structures-a technique developed at chemical abstracts service.''
  \emph{Journal of Chemical Documentation}, vol.~5, no.~2, pp. 107--113, 1965.

\bibitem{durant2002reoptimization}
J.~L. Durant, B.~A. Leland, D.~R. Henry, and J.~G. Nourse, ``Reoptimization of
  mdl keys for use in drug discovery,'' \emph{Journal of chemical information
  and computer sciences}, vol.~42, no.~6, pp. 1273--1280, 2002.

\bibitem{landrum2013rdkit}
G.~Landrum, ``Rdkit: A software suite for cheminformatics, computational
  chemistry, and predictive modeling,'' 2013.

\bibitem{real1996probabilistic}
R.~Real and J.~M. Vargas, ``The probabilistic basis of jaccard's index of
  similarity,'' \emph{Systematic biology}, vol.~45, no.~3, pp. 380--385, 1996.

\bibitem{JDH17}
J.~Johnson, M.~Douze, and H.~J{\'e}gou, ``Billion-scale similarity search with
  gpus,'' \emph{arXiv preprint arXiv:1702.08734}, 2017.

\bibitem{faiss}
H.~Jégou, M.~Douze, J.~Johnson, and L.~Hosseini, ``Faiss,''
  \url{https://github.com/facebookresearch/faiss}, 2021.

\bibitem{boytsov2013sisap}
\BIBentryALTinterwordspacing
L.~Boytsov and B.~Naidan, ``Engineering efficient and effective non-metric
  space library,'' in \emph{Similarity Search and Applications - 6th
  International Conference, {SISAP} 2013, {A} Coru{\~{n}}a, Spain, October 2-4,
  2013, Proceedings}, ser. Lecture Notes in Computer Science, N.~R. Brisaboa,
  O.~Pedreira, and P.~Zezula, Eds., vol. 8199.\hskip 1em plus 0.5em minus
  0.4em\relax Springer, 2013, pp. 280--293. [Online]. Available:
  \url{https://doi.org/10.1007/978-3-642-41062-8\_28}
\BIBentrySTDinterwordspacing

\bibitem{NMSLIB}
L.~Boytsov \emph{et~al.}, ``Non-metric space library (nmslib),''
  \url{https://github.com/nmslib/nmslib}, 2021.

\bibitem{andoni2015practical}
A.~Andoni, P.~Indyk, T.~Laarhoven, I.~Razenshteyn, and L.~Schmidt, ``Practical
  and optimal lsh for angular distance,'' \emph{arXiv preprint
  arXiv:1509.02897}, 2015.

\bibitem{FALCONN}
I.~Razenshteyn and L.~Schmidt, ``Falconn - fast lookups of cosine and other
  nearest neighbors,'' \url{https://github.com/FALCONN-LIB/FALCONN}, 2017.

\bibitem{zhu2020benchmark}
C.~J. Zhu, M.~Song, Q.~Liu, C.~Becquey, and J.~Bi, ``Benchmark on indexing
  algorithms for accelerating molecular similarity search,'' \emph{Journal of
  Chemical Information and Modeling}, 2020.

\bibitem{swamidass2007bounds}
S.~J. Swamidass and P.~Baldi, ``Bounds and algorithms for fast exact searches
  of chemical fingerprints in linear and sublinear time,'' \emph{Journal of
  chemical information and modeling}, vol.~47, no.~2, pp. 302--317, 2007.

\bibitem{nasr2010hashing}
R.~Nasr, D.~S. Hirschberg, and P.~Baldi, ``Hashing algorithms and data
  structures for rapid searches of fingerprint vectors,'' \emph{Journal of
  chemical information and modeling}, vol.~50, no.~8, pp. 1358--1368, 2010.

\bibitem{malkov2018efficient}
Y.~A. Malkov and D.~A. Yashunin, ``Efficient and robust approximate nearest
  neighbor search using hierarchical navigable small world graphs,'' \emph{IEEE
  transactions on pattern analysis and machine intelligence}, vol.~42, no.~4,
  pp. 824--836, 2018.

\bibitem{mendez2019chembl}
D.~Mendez, A.~Gaulton, A.~P. Bento, J.~Chambers, M.~De~Veij, E.~F{\'e}lix,
  M.~P. Magari{\~n}os, J.~F. Mosquera, P.~Mutowo, M.~Nowotka \emph{et~al.},
  ``Chembl: towards direct deposition of bioassay data,'' \emph{Nucleic acids
  research}, vol.~47, no.~D1, pp. D930--D940, 2019.

\bibitem{HNSWLIB}
e.~a. Malkov, ``Hnswlib,'' \url{https://github.com/nmslib/hnswlib}, 2021.

\bibitem{danopoulos2019approximate}
D.~Danopoulos, C.~Kachris, and D.~Soudris, ``Approximate similarity search with
  faiss framework using fpgas on the cloud,'' in \emph{International Conference
  on Embedded Computer Systems}.\hskip 1em plus 0.5em minus 0.4em\relax
  Springer, 2019, pp. 373--386.

\end{thebibliography}
}



\end{document}